\documentclass[a4j]{article}
\pdfoutput=1

\usepackage[dvipdfmx]{graphicx}
\usepackage{mediabb}
\usepackage{amsmath}
\usepackage{amssymb}
\usepackage{mathtools}
\usepackage{authblk}
\usepackage{xcolor}
\usepackage{here}
\usepackage{listings}
\usepackage{algorithm}
\usepackage{algorithmicx}
\usepackage{algpseudocode}
\usepackage[margin=1.5in]{geometry}

\usepackage{cite}
\usepackage[hang,small]{caption}
\usepackage[subrefformat=parens]{subcaption}
\def\*#1{\boldsymbol{#1}}
\newcommand{\tw}[0]{\textwidth}

\newcommand{\igr}[2]{\includegraphics[clip,width=#1\tw]{#2}}

\newcommand{\eq}[1]{(\ref{#1})}

\newcommand{\lw}[1]{\smash{\lower2.ex\hbox{#1}}}

\newcommand{\rbr}[1]{\left(#1\right)}

\newcommand{\RR}{\mathbb{R}}

\newcommand{\EE}{\mathbb{E}}

\newcommand{\cD}{{\cal D}}

\newcommand{\cT}{{\cal T}}

\title{
Cost-effective search for lower-error region in \\
material parameter space using multifidelity \\
Gaussian process modeling
}
\author[1]{Shion Takeno}
\author[2,3]{Yuhki Tsukada\thanks{Corresponding author}}
\author[2]{Hitoshi Fukuoka}
\author[2]{Toshiyuki Koyama}
\author[4,3,5]{Motoki Shiga}
\author[1,3,6]{Masayuki Karasuyama$^*$}

\affil[1]{Department of Computer Science, Graduate School of Engineering, Nagoya Institute of Technology, Gokiso-cho, Showa-ku, Nagoya, 466-8555, Japan}
\affil[2]{Department of Materials Design Innovation Engineering, Graduate School of Engineering, Nagoya University, Furocho, Chikusa-ku, Nagoya, 464-8603, Japan}
\affil[3]{JST, PRESTO, 4-1-8 Honcho, Kawaguchi, Saitama, 332-0012, Japan}
\affil[4]{Department of Electrical, Electronic and Computer Engineering, Faculty of Engineering, Gifu University, 1-1 Yanagido, Gifu, 501-1193, Japan}
\affil[5]{Center for Advanced Intelligence Project, RIKEN, 1-4-1 Nihonbashi, Chuo-ku, Tokyo, 103-0027, Japan}
\affil[6]{Center for Materials Research by Information Integration, National Institute for Materials Science, 1-2-1 Sengen, Tsukuba, Ibaraki, 305-0047, Japan}

\date{}

\newcommand{\comment}[1]{}

\begin{document}

\maketitle

\clearpage

\begin{abstract}
Information regarding precipitate shapes is critical for estimating material parameters.
Hence, we considered estimating a region of material parameter space in which a computational model produces precipitates having shapes similar to those observed in the experimental images.
This region, called the lower-error region (LER), reflects intrinsic information of the material contained in the precipitate shapes.
However, the computational cost of LER estimation can be high because the accurate computation of the model is required many times to better explore parameters.
To overcome this difficulty, we used a Gaussian-process-based multifidelity modeling, in which training data can be sampled from multiple computations with different accuracy levels (fidelity).
Lower-fidelity samples may have lower accuracy, but the computational cost is lower than that for higher-fidelity samples.
Our proposed sampling procedure iteratively determines the most cost-effective pair of a point and a fidelity level for enhancing the accuracy of LER estimation.
We demonstrated the efficiency of our method through estimation of the interface energy and lattice mismatch between MgZn$_2$ and $\alpha$-Mg phases in an Mg-based alloy.
The results showed that the sampling cost required to obtain accurate LER estimation could be drastically reduced.
\end{abstract}

\clearpage

% --------------------------------------------------
\section{Introduction}
Material parameters are often estimated by fitting a theory or model to experimentally observed microstructures.
For example, the interface energy between precipitate and matrix phases is estimated by fitting the Ostwald ripening model \cite{Kahlweit1975} (theoretical formula) to time-series experimental data of the precipitate radius during the coarsening process.
Some recent studies estimated material parameters by comparing data regarding microstructure evolution obtained through experiments and simulations \cite{Ito2016,Ito2017, Zhang2017,Sasaki2018}.
Because a precipitate prefers an energetically favorable shape \cite{Thompson1994,Schmidt1997,Schmidt1998,Khachaturyan2008,Porter2009}, information about precipitate shapes is valuable for estimating material parameters.
In Mg-based alloys, rod- or plate-shaped precipitates with various aspect ratios have been observed \cite{Clark1965,Chun1969,Nie1997,Celotto2000,Smola2002,Ping2003,Oh2005,Nie2005,Sasaki2006,Mendis2006,Mendis2007,Sasaki2009,Oh-ishi2009,Sasaki2011,Mendis2011,Mendis2012,Elsayed2013,Bhattacharjee2013,Bhattacharjee2014,Sasaki2015,Nakata2017}.
Moreover, precipitate shapes can be predicted using some advanced computational models if the interface energy and lattice mismatch between the precipitate and matrix phases are given \cite{Gao2012,Liu2013,Ji2014,Tsukada2014}.
Hence, fitting the computational models to experimental data on precipitate shape enables us to estimate material parameters.
However, parameter estimation based on precipitate shapes is time-consuming because the computational cost for predicting precipitate shapes is high.

\begin{figure}[t]
 \centering
 \igr{0.7}{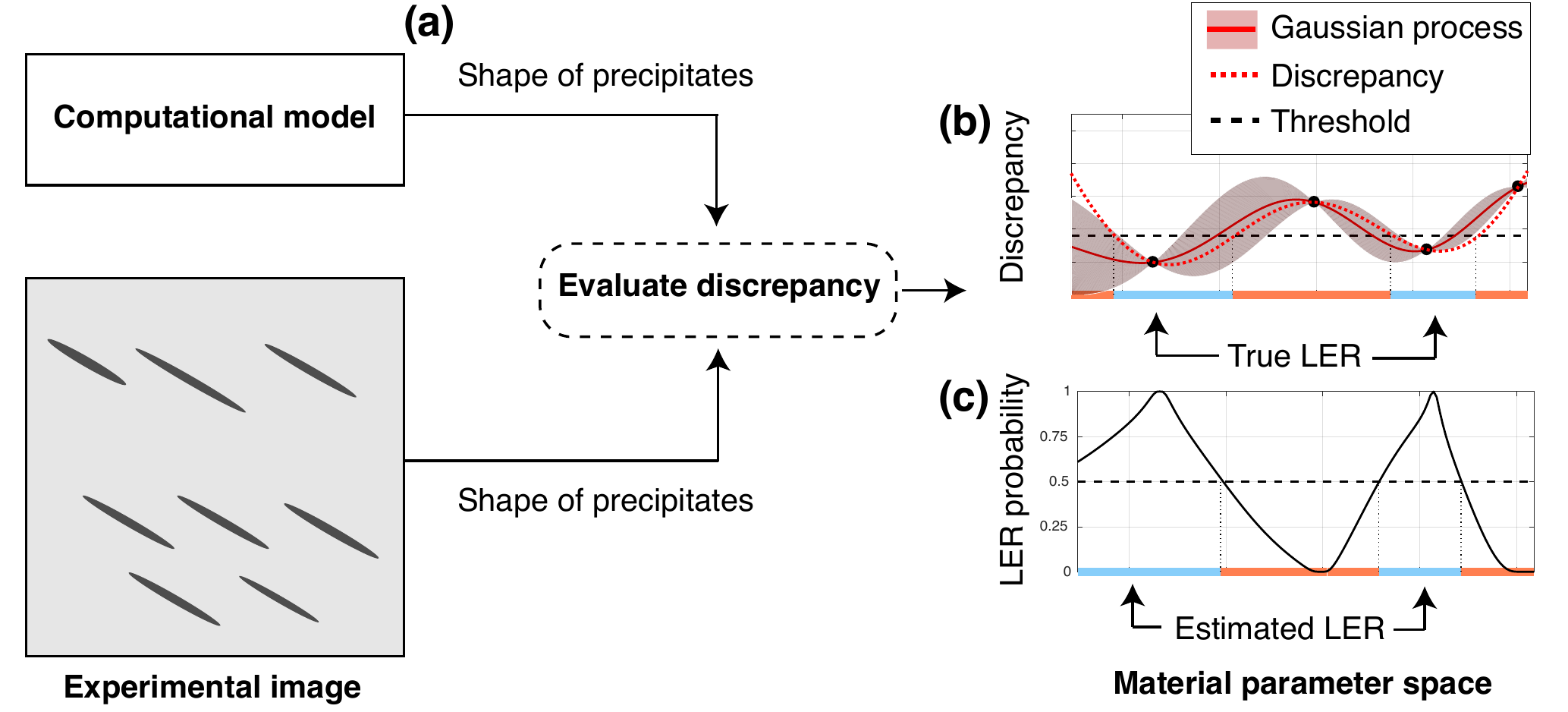}
 \caption{
 LER estimation by GP.
 (a) Discrepancy between the computational model and experimental image is evaluated through the difference in the precipitate shapes.
 (b) Using few observed discrepancy values (black circles) in the material parameter space, GP regression approximates discrepancy surface.
 The red dotted line is the underlying true discrepancy that is unknown beforehand.
 The solid line and shaded regions represent the GP regression and its predictive variance, respectively.
 (c) Probability of LER estimated by the GP model.
 From the GP, the probability that each material parameter has a discrepancy value smaller than the threshold can be estimated.
 If the probability is more than 0.5, the region is estimated as LER.
 }
 \label{fig:LSE}
\end{figure}

Therefore, to mitigate this problem, we recently introduced a \emph{Gaussian process} (GP)-based selective sampling procedure for material parameter estimation from precipitate shapes \cite{Tsukada2019-Estimation}.
Figure~\ref{fig:LSE} shows a schematic illustration of this approach.
When we have a computational model that predicts the energetically favorable shape of the precipitate under given material parameters, we can calculate the discrepancy between the precipitate shape observed in the experiment and that predicted using the computational model.
Because experimental data on precipitate shapes are naturally uncertain, the exact minimum of the discrepancy is not necessarily a unique optimal parameter.
Instead, the \emph{lower-error region} (LER) of the material parameter space, in which the discrepancy is smaller than a given threshold, is estimated.
By determining the threshold from the variance of the precipitate shapes in the experiment, LER estimation can provide a region with reasonable parameters that can be consistent with the current experimental result.

% --------------------------------------------------
% Illustration of LER
% --------------------------------------------------

Although GP-based LER identification can be much more efficient than the exhaustive search or na{\" i}ve random sampling methods, obtaining accurate shapes of the precipitates at every iteration requires a considerably high computational cost.
However, by controlling the accuracy of numerical computations, we can also obtain approximate discrepancy values with much lower computational costs.
In a computational model for predicting precipitate shapes \cite{Tsukada2014}, the total energy (sum of strain and interface energies) of a spheroidal precipitate is formulated as a function of the precipitate aspect ratio $r$ if the material parameters are given.
By computing the total energy using different values of $r$, the equilibrium shape (aspect ratio) of the precipitate that minimizes the total energy can be predicted.
If we change the step size of $r$ in the numerical computation, the tradeoff between the computational cost and accuracy can be controlled.
In this study, we considered GP-based LER estimation that adaptively incorporates training data from different levels of approximate calculations.
The degree of approximation is called {\it fidelity}.
Although lower-fidelity data contain stronger approximations, it is often useful to narrow the candidate region during early-stage screening in our material parameter exploration.
We considered efficiently identifying LER by sampling discrepancies not only from the highest-fidelity calculations but also from lower-fidelity calculations that are much easier to perform.

\begin{figure}[t]
 \centering
 \igr{0.95}{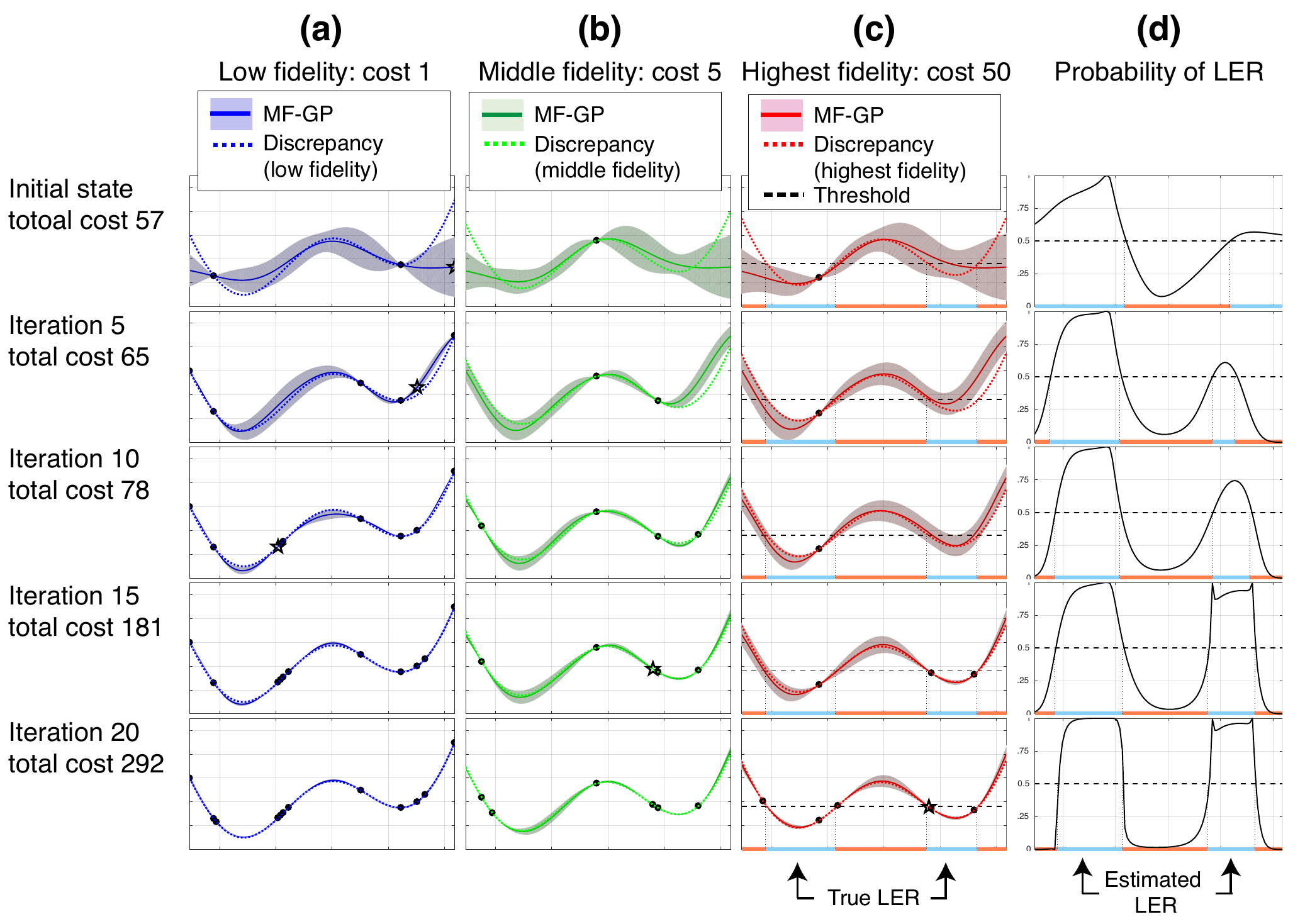}
 \caption{
 Illustrative example of proposed multifidelity LER estimation.
 The three columns on the left,(a)--(c) show the low-, middle-, and highest-fidelity functions having different sampling costs (1, 5, and 50), and (d) shows the probability of LER estimation from the GP for the highest fidelity.
 The three different fidelity functions are fitted using the multifidelity GP (MF-GP) in which information of the sampled points from multiple fidelities are shared with each other.
 The star point is the next candidate determined by our sampling criterion.
 At the beginning of the iterations, the low-fidelity function is mainly sampled.
 This is because GP prediction is highly uncertain in the early steps, and then, large amount of information can be obtained even from the low-fidelity function.
 At iteration 10 (the third row), the probability of LER estimation shown in (d) roughly corresponds to LER without any additional sample in the highest-fidelity function.
 At iteration 20 (the bottom row), by sampling from the highest-fidelity function, a more accurate LER estimation is obtained, though the number of samples in the highest-fidelity function is still only four.
 }
 \label{fig:toy-demo-mflse}
\end{figure}

{\it Multifidelity modeling} is a machine-learning (ML) framework that combines inexpensive lower-accuracy data and expensive higher-accuracy data to estimate a model with a lower sampling cost of the training data.
Figure~\ref{fig:toy-demo-mflse} shows an illustrative example of our proposed multifidelity LER estimation procedure.
%
% In the figure, % GP integrates different fidelity samples by which the highest fidelity function can be inferred through information from the low- and middle-fidelity functions.
As shown in the figure, GP integrates different fidelity samples through which similarities among different fidelity functions are automatically estimated, and information from the low- and middle-fidelity functions enhance the inference of the highest-fidelity function.
% the highest fidelity function can be inferred through
% information from the low and middle fidelity functions
%
Our cost-effective sampling criterion is based on {\it information entropy}, which evaluates the uncertainty of the probabilistic estimation.
At every iteration, the most cost-effective pair of a sampling point and a fidelity level can be selected for reducing the uncertainty of LER in terms of information entropy.
%
% As Fig.~\ref{fig:toy-demo-mflse} illustrates, this enables us to estimate LER efficiently by (1) sampling only a small number of points compared with the exhaustive search, and (2) fully utilizing lower fidelity information.
As Fig.~\ref{fig:toy-demo-mflse} illustrates, this method enables us to estimate LER efficiently by sampling only a small number of points compared with the exhaustive search; in particular, we can avoid sampling of higher-fidelity functions many times that results in high computational costs.
Although multifidelity modeling is used in materials science applications such as bandgap predictions \cite{Pilania2017-Multi}, to the best of our knowledge, our study is the first of its kind using a multifidelity-based exploration algorithm involving material parameters.
We applied our proposed method called multifidelity LER (MF-LER) estimation to estimate the interface energy and lattice mismatch between MgZn$_2$ ($\beta_1^\prime$) and $\alpha$-Mg phases in an Mg-based alloy, in which we have three different fidelity levels requiring 5, 10, and 60 minutes to compute, respectively.
We demonstrated that our approach drastically accelerated the material parameter search by efficiently using lower-fidelity samples.
Although we focused on an Mg-based alloy in our study, MF-LER is applicable to other material parameter estimation problems because multifidelity calculations are prevalent in computational materials science, in which the computational cost often becomes a severe bottleneck.

% --------------------------------------------------
% Demo MF-LSE
% --------------------------------------------------

\clearpage
% --------------------------------------------------
\section{Methods}

% -------------------------
\subsection{Problem Setting}

Let $r_{\rm expt}$ be the aspect ratio of the precipitate obtained from an experimental image, and $r^{(m)}_{\*x_i, {\rm comput}}$ be the aspect ratio predicted using a computational model with the material parameter $\*x_i \in \RR^d$ and the fidelity level $m \in \{1, \ldots, M\}$.
We assumed a set of $N$ candidates $\{ \*x_i \}_{i = 1}^N$ in the material parameter space (for example, grid points uniformly taken in the space).
If the higher-fidelity level $m$ is calculated, more accurate results can be obtained though it requires a higher computational cost.
Let
$\lambda^{(1)} \leq \lambda^{(2)} \leq \ldots \leq \lambda^{(M)}$
be the sampling cost of each fidelity (computational time of the model).
The discrepancy between the aspect ratios obtained from the experimental image and through the computational model is defined by
\begin{align*}
 y^{(m)}_{\*x_i}
 =
 \frac{1}{2} \sum_{t \in \cT}
 (r_{\rm expt}(t) - r^{(m)}_{\*x_i, {\rm comput}}(t))^2,
\end{align*}
where $t$ is the time and $\cT$ is a set of times when the shapes of precipitates are experimentally measured.

Suppose that the observed discrepancy contains an independent additive noise term given as follows:
$y^{(m)}_{\*x} = f^{(m)}_{\*x} + \epsilon$,
where
$\epsilon \sim \mathcal{N}(0, \sigma_{\rm noise}^2)$.
Then, the LER, in which the true discrepancy of the highest-fidelity function $f^{(M)}_{\*x}$ is less than a given threshold $h$, is defined as
\begin{align*}
 LER = \left\{ i \mid f_{\*x_i}^{(M)} \leq h \right\}.
\end{align*}
If a large set of the highest-fidelity values of $y_{\*x_i}^{(M)}$ can be obtained for a variety of $\*x_i$, LER can be identified accurately.
However, this leads to prohibitive computational costs because the fidelity level $M$ needs the highest computational cost, and further, the number of candidate material parameters is often high.
%
% LER is usually only a small part of the material parameter space
%
Our goal is to identify the LER with the small total sampling cost (the sum of $\lambda^{(m)}$ over the sampled points).

% -------------------------
\subsection{Multifidelity Gaussian Process}

Suppose we already have the dataset
$\cD_{n} = \{ (\*x_i, y^{(m_i)}_{\*x_i}, m_i) \}_{i=1}^n$
containing a set of triplets consisting of an input $\*x_i \in \RR^d$, fidelity $m_i \in \{ 1, \ldots, M \}$ and output $y^{(m_i)}_{\*x_i} \in \RR$.
To jointly model different fidelity observations with the GP, we used a multifidelity extension of GP regression (MF-GP) \cite{Kennedy2000-Predicting}, which is also known as a co-kriging model.
Let $f^{(M)} \sim GP(0, k_M(\*x,\*x'))$ be the GP for the highest fidelity $m = M$, in which the prior mean is $0$ and the covariance function is $k_M: \RR^d \times \RR^d \rightarrow \RR$ (the covariance function is also called kernel function).
Note that we can set the prior mean as $0$ without loss of generality.
We define the output for the lower fidelity $m = M - 1, \ldots, 1$ recursively from $M$ as follows;
\begin{align*}
 f^{(m)}_{\*x} & = f^{(m+1)}_{\*x} + g^{(m+1)}_{\*x},
 % f^{(m)}_{\*x} & = f^{(m-1)}_{\*x} + g^{(m-1)}_{\*x},
 % \label{eq:fidelity-diff}
\end{align*}
where
$g^{(m+1)} \sim GP(0, k_g(\*x,\*x^\prime))$
with the kernel function
$k_g: \RR^d \times \RR^d \rightarrow \RR$.
The function $g^{(m+1)}_{\*x}$ represents difference between
$f^{(m)}_{\*x}$
and
$f^{(m+1)}_{\*x}$.
For example, when $M = 3$, we obtain
$f^{(2)}_{\*x}  = f^{(3)}_{\*x} + g^{(3)}_{\*x}$, and
% $f^{(3)}_{\*x}  = f^{(2)}_{\*x} + g^{(2)}_{\*x}$.
$f^{(1)}_{\*x} = f^{(2)}_{\*x} + g^{(2)}_{\*x} = f^{(3)}_{\*x} + g^{(3)}_{\*x}+ g^{(2)}_{\*x}$.
% Therefore, $f^{(m)}_{\*x}$ and $f^{(m')}_{\*x}$ are more likely to be similar probabilistically, when $m$ and $m'$ are close each other.
%
% Therefore, the difference between different fidelity functions in this model.
The difference between $f^{(3)}_{\*x}$ and $f^{(2)}_{\*x}$, which have neighboring fidelity levels, is modeled using the single GP model $g^{(3)}_{\*x}$.
In contrast, the difference between $f^{(3)}_{\*x}$ and $f^{(1)}_{\*x}$, whose fidelity levels are more distant from each other, is modeled by the sum of the two GP models $g^{(3)}_{\*x}$ and $g^{(2)}_{\*x}$.
As a result, in this model, the difference between
$f^{(3)}_{\*x}$
and
$f^{(1)}_{\*x}$
has a larger variance, compared with
$f^{(2)}_{\*x}$
and
$f^{(1)}_{\*x}$.

In MF-GP, the kernel function for a pair of training instances
$\{ (\*x_i,y^{(m_i)}_{\*x_i},m_i), (\*x_j,y^{(m_j)}_{\*x_j},m_j) \}$
is written as
$k((\*x_i,m_i),(\*x_j,m_j)) = k_1(\*x_i,\*x_j) + ( \min(m_i, m_j)-1) k_g(\*x_i,\*x_j)$
(see \cite{Kennedy2000-Predicting} for detail).
Using the kernel matrix $\*K \in \RR^{n \times n}$ in which element $i,j$ is defined by
$k((\*x_i,m_i),(\*x_j,m_j))$,
the GP for all fidelities $f^{(1)}, \ldots, f^{(M)}$ can be integrated into one GP in which the predictive mean and variance are obtained as
\begin{align*}
 \mu^{(m)}_{\*x}
 & =
 \*k^{(m)}_n(\*x)^\top
 \rbr{\*K + \sigma_{\rm noise}^2 \*I}^{-1}
 \*y,
 % \label{eq:MF-GPR-mean}
 \\
 \sigma^{2(m)}_{\*x}
 & =
 k((\*x,m),(\*x,m))
 \nonumber \\
 & - \*k^{(m)}_n(\*x)^\top
 \rbr{\*K + \sigma_{\rm noise}^2 \*I}^{-1}
 \*k^{(m)}_n(\*x),
 % \label{eq:MF-GPR-var}
\end{align*}
where
$\*y = (y^{(m_1)}_{\*x_1}, \ldots, y^{(m_n)}_{\*x_n})^\top$
and
$\*k^{(m)}_n(\*x) \coloneqq (k((\*x,m),(\*x_1,m_1)), \ldots, k((\*x,m),(\*x_n,m_n)))^\top$.
Each row of Fig.~\ref{fig:toy-demo-mflse} (a)--(c) shows examples of MF-GP for $M = 3$.
As shown in the figure, information from the training data is transferred across different fidelities.
As a result, the similarities among different fidelity functions are automatically estimated, and
% the information from lower fidelity functions enhance
the inference of the highest-fidelity function is enhanced by the lower-fidelity observations.

% -------------------------
\subsection{Sampling criterion for LER estimation}

Estimating the LER can be considered a classification problem in which each input $\*x_i$ is classified based on whether it is included in $LER$.
Let
\begin{align*}
z_{\*x} &=
 \begin{cases}
  1, & \text{ if } f^{(M)}_{\*x} \leq h, \\
  0, & \text{ if } f^{(M)}_{\*x} > h,
 \end{cases}
\end{align*}
be an indicator variable of the LER classification.
From the definition, we obtain
$p(z_{\*x} = 1) = p(f^{(M)}_{\*x} \leq h)$ and
$p(z_{\*x} = 0) = p(f^{(M)}_{\*x} > h)$.
If
$p(z_{\*x} = 0)$
and
$p(z_{\*x} = 1)$
are considerably different, the confidence of the prediction is considered high, while
%; however,
if these two values are close to $0.5$, the confidence of the prediction is considered low.

For the dataset
$\cD_{n} = \{ (\*x_i, y^{(m_i)}_{\*x_i}, m_i) \}_{i=1}^n$,
the total cost of sampling is $\sum_{i = 1}^n \lambda^{(m_i)}$.
We considered estimating accurate $z$ with the small total sampling cost.
To evaluate the benefit of sampling from a variety of fidelity levels, we used {\it information theory} \cite{MacKay2003-Information}.
%  which provides a measure of uncertainty of the probabilistic prediction.
%
Let
$p(z_{\*x} | \mathcal{D}_n)$ be the conditional distribution of $z_{\*x}$, given the training data $\mathcal{D}_n$, and
$p(z_{\*x} | y^{(m)}_{\*x}, \mathcal{D}_n)$ be the conditional distribution of $z_{\*x}$, given the training data $\mathcal{D}_n$, and a new observation $y^{(m)}_{\*x}$.
For a pair of $m$ and $\*x$, the amount of information, called {\it information gain}, obtained about $z_{\*x}$ through observing $y^{(m)}_{\*x}$ is written as
\begin{align}
 I(z_{\*x};y^{(m)}_{\*x}) =&
 H(p(z_{\*x} | \mathcal{D}_n)) -
 \mathbb{E}_{p(y^{(m)}_{\*x} | \mathcal{D}_n)}
 \bigl[
 H(p(z_{\*x} | y^{(m)}_{\*x}, \mathcal{D}_n))
 \bigl],
\label{eq:mutual_info}
\end{align}
where
$H$ is the information entropy and
$\EE_{p(y^{(m)}_{\*x} | \cD_t)}$
is the expectation over $y^{(m)}_{\*x}$.
%
% We call $I(z_{\*x};y^{(m)}_{\*x})$ in \eq{eq:mutual_info} {\it information gain}.
%
Information entropy $H$ is a standard uncertainty measure of a random variable in information theory,
which is defined as
$H[p(X)] = \EE_{p(X)}[-\log_2(p(X))]$
for a random variable $X$.
In our case, both
$p(z_{\*x} | \mathcal{D}_n)$
and
$p(z_{\*x} | y^{(m)}_{\*x}, \mathcal{D}_n)$
are Bernoulli distributions.
%
% For our Bernoulli distribution $p(z)$ ($z \in \{0,1\}$), information entropy is
% $H(p(z)) = - p(z) \log p(z) - (1-p(z)) \log (1-p(z))$
In general, for a Bernoulli distribution $p(z)$ ($z \in \{0,1\}$), information entropy is
$H(p(z)) = - p(1) \log_2 p(1) - p(0) \log_2 p(0)$
%Information entropy $H$ is an uncertainty measure of a random variable
that takes the maximum value $1$ when
% $p(z = 1) = p(z = 0) = 1/2$
$p(1) = p(0) = 1/2$ (most uncertain)
and the minimum value $0$ when
$p(1) = 0$ and $p(0) = 1$,
or
$p(1) = 1$ and $p(0) = 0$.
Thus, the first term of \eq{eq:mutual_info} is the uncertainty of the current $z_{\*x}$, and the second term is the expected uncertainty after adding the candidate $y^{(m)}_{\*x}$ into the training data.
Because $y^{(m)}_{\*x}$ is not observed yet,
the expectation is taken over the current GP estimation of
$p(y^{(m)}_{\*x} | \mathcal{D}_n)$.
In other words, \eq{eq:mutual_info} can be seen as the expected uncertainty reduction after sampling $y_{\*x}^{(m)}$.
Computational details of the information gain are provided in supplementary appendix~\ref{sec:computation-entropy}.

\begin{figure}
 \centering
 \igr{0.9}{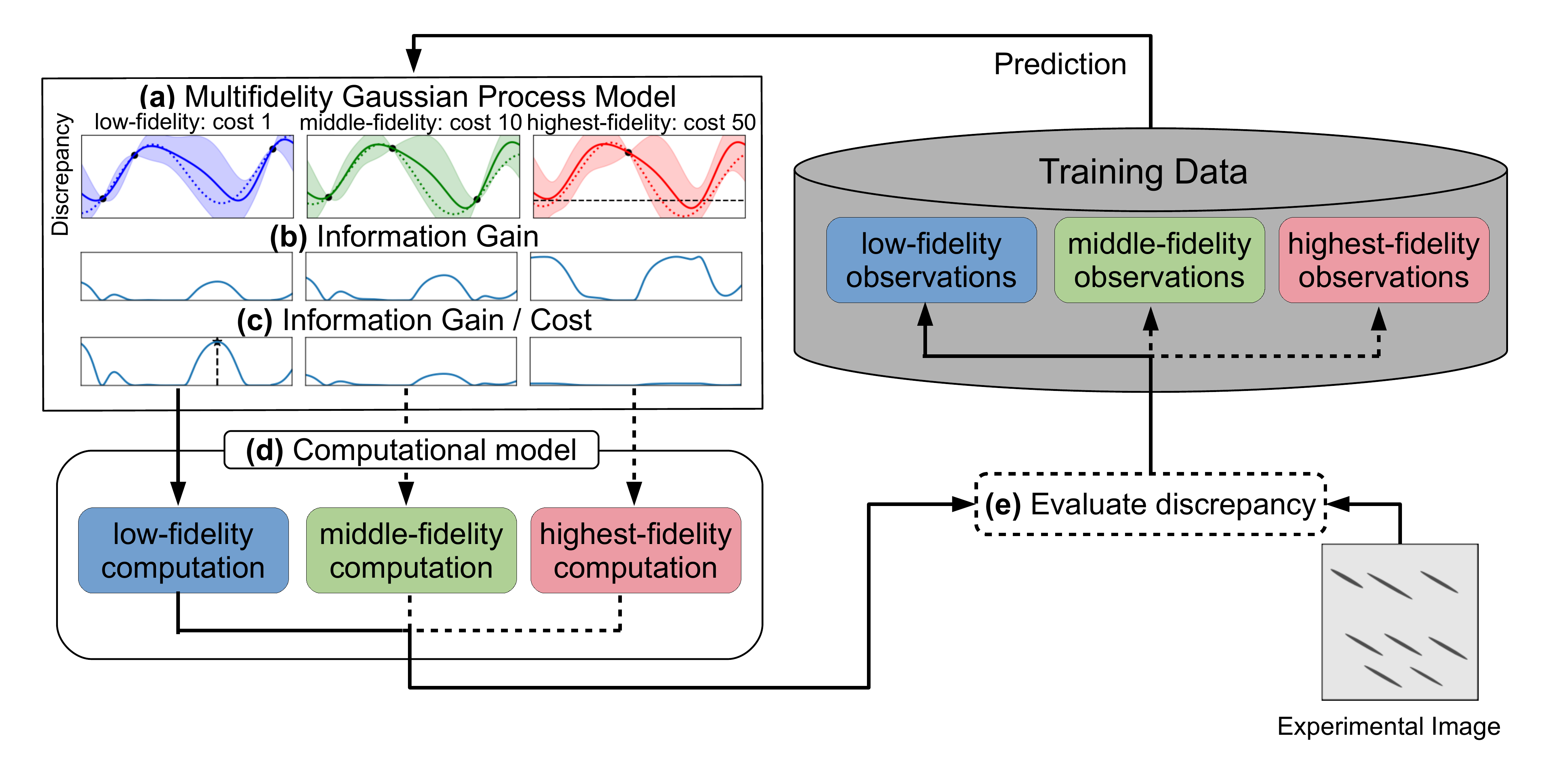}
 \caption{
 Schematic illustration of MF-LER.
 (a) MF-GP provides predictions using all the observations across different fidelities.
 The cost values for the low-, middle-, and highest- fidelity functions in this illustration are 1, 10, and 50, respectively.
 (b) Information gain for identifying the LER in the highest-fidelity function is evaluated through the MF-GP model.
 (c) Information gain is divided by the sampling cost, which enables us to evaluate the cost-effectiveness of sampling.
 (d) The computational model is calculated with the selected fidelity and material parameters (in this illustration, the low-fidelity function is selected).
 (e) Discrepancy between the computational model and experimental image is evaluated (through the precipitate shapes), and the result is added to the training data.
 }
 \label{fig:procedure}
\end{figure}

For higher $m$, a
larger amount of information about $z_{\*x}$ can be obtained.
However, this requires a higher sampling cost. %, and thus, cost-effectiveness of each candidate should be considered to determine a next sampling point.
We selected a pair of $m$ and $\*x_i$ that maximizes the following cost-effectiveness criterion $a(\*x_i, m)$, in which information gain is divided by the sampling cost of $y^{(m)}_{\*x_i}$:
\begin{align*}
 a(\*x_i, m) = \frac{I(z_{\*x_i}; y^{(m)}_{\*x_i}) }{\lambda^{(m)}}.
\end{align*}
%
% We call this criterion acquisition function.
%
% Figure~\ref{fig:acq_func} shows a schematic illustration of our sampling criterion.
%
Because this criterion represents the amount of information per unit sampling cost, our sampling process can be efficient in terms of the actual computational cost rather than the number of iterations.
Figure~\ref{fig:procedure} shows the entire procedure of our method, called MF-LER (multifidelity LER estimation), in which the most cost-effective pair of a sampling point and a fidelity level is iteratively selected.
Further, a demonstration using a simple one-dimensional function is shown in Fig.~\ref{fig:toy-demo-mflse}.
We can see that the lower-fidelity functions are fully utilized for identifying LER efficiently.

% --------------------------------------------------
% Procedure
% --------------------------------------------------

\clearpage
% --------------------------------------------------
\section{Results}

% -------------------------
\subsection{Computational Model for Predicting Precipitate Shape}

We assumed a rod- or plate-shaped precipitate as a spheroid ($x^2 / a^2 + y^2 / b^2 + z^2 / c^2 = 1, a = b, r = c / a$).
The total energy (sum of strain energy and interface energy) of the spheroidal precipitate is formulated as
\begin{align*}
	E_{\mathrm{total}}(r) &\equiv E_{\mathrm{strain}}(r) + E_{\mathrm{interface}}(r) \\
	&= \frac{V_0}{2} C_{ijkl}\varepsilon^0_{kl} \bigl(\varepsilon^0_{ij} - S_{ijmn}(r)\varepsilon^0_{mn} \bigl) + A(r)\gamma,
\end{align*}
where $V_0$ is the precipitate volume, $C_{ijkl}$ is the elastic modulus tensor, $\varepsilon^0_{ij}$ is the crystal lattice mismatch between the precipitate and matrix phases, $S_{ijmn}(r)$ is Eshelby's tensor \cite{Mura1987}, $A(r)$ is the interface area, and $\gamma$ is the interface energy (in $\mathrm{J}\mathrm{m}^{-2}$).
The formulas for computing $S_{ijmn}(r)$ and $A(r)$ can be found in ref. \cite{Tsukada2014}.
When values of the material parameters are given, we can compute $E_{\mathrm{total}}$ as a function of the aspect ratio of the spheroid $r$, which is changed from $1$ to $100$ with a step size $\Delta r$.
Then, the aspect ratio that minimizes $E_{\mathrm{total}}$ is $r_{\mathrm{comput}}$ that corresponds to the equilibrium precipitate shape.
In this study, we computed the aspect ratio of the $\mathrm{MgZn}_2$ phase in the $\alpha$-Mg phase. The $\varepsilon^0_{ij}$ is given by
\begin{align*}
	\varepsilon^0_{ij} = \left( \begin{array}{ccc}
		\varepsilon^0_{11} & 0 & 0 \\
		0 & \varepsilon^0_{22} & 0 \\
		0 & 0 & 0.00182
	\end{array} \right),
\end{align*}
where $\varepsilon^0_{11} = \varepsilon^0_{22}$ \cite{Tsukada2014,Tsukada2019-Estimation}.
The elastic modulus tensor for the $\alpha$-Mg phase \cite{Kinzoku2004} was used for the computation.
Computational models with low-, middle- and highest-fidelity functions were prepared by setting $\Delta r$ as $10^{-3}, 10^{-4}$ and $10^{-5}$,  respectively.
We considered estimating the interface energy $\gamma$ and lattice mismatch $\varepsilon^0_{11}$ between the $\mathrm{MgZn}_2$  and $\alpha$-Mg phases using the experimental data on the changes in the aspect ratio of the rod-shaped $\mathrm{MgZn}_2$ phase in an Mg-based alloy aged at $160~ ^\circ \mathrm{C}$ for $2, 8$, and $24$ hours \cite{Bhattacharjee2013}.

% -------------------------
\subsection{Performance Evaluation}

% --------------------------------------------------
% Surface
% --------------------------------------------------
\begin{figure}[t]
 \centering
 \igr{0.5}{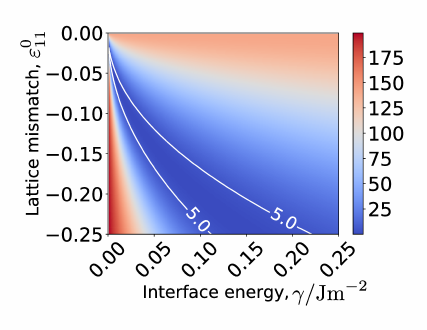}
 \caption{
 Heatmap of highest fidelity discrepancy and LER defined by threshold $h = 5$.
 }
 \label{fig:surface}
\end{figure}

In our study, we demonstrated the performance of MF-LER using the Mg-based alloy data.
On the basis of the analysis in our previous study \cite{Tsukada2019-Estimation}, we set $h = 5$ that was empirically inferred from the standard deviation of the aspect ratio in the experimental image.
Figure~\ref{fig:surface} shows the heatmap of discrepancy and LER.
We have three fidelity levels $M = 3$, and the sampling costs of the low-, middle-, and highest-fidelity functions are $\lambda^{(1)} = 5$, $\lambda^{(2)} = 10$, and $\lambda^{(3)} = 60$ minutes, respectively.
For the candidate parameter $\*x = (\gamma, \varepsilon_{11}^0)^\top$, we used 250 equally spaced grids in
$\varepsilon_{11}^0 \in [-0.250, -0.001]$
and
$\gamma \in [0.001, 0.250]$ (J m$^{-2}$).
Thus, we have a total of $N = 62500$ candidates that require $3750000 (= 62500 \times 60)$ minutes to compute all the points in the highest-fidelity function.
To evaluate the usefulness of the low-fidelity observations, we compared MF-LER with two strategies that take samples only from the highest-fidelity function.
The first approach is to use information gain \eq{eq:mutual_info} as the sampling criterion \cite{Tsukada2019-Estimation}, called single-fidelity LER estimation (SF-LER), and the second approach is single-fidelity GP with random sampling, called SF-Random.
For the initial points, SF-Random and SF-LER randomly selected five highest-fidelity points, and MF-LER randomly selected ten lowest fidelity points.
For all approaches, a candidate $\*x$ is classified as LER if $p(z_{\*x} = 1) \geq 0.5$. % , which is equivalent to $\mu^{(M)}_{\*x} \leq h$.
Detailed explanations of the GP are given in supplementary appendix~\ref{sec:settings}.

\figurename~\ref{fig:demo} shows the sampling processes of MF-LER, SF-LER, and SF-Random.
At ``Cost 500,'' MF-LER determines LER approximately, while SF-LER and SF-Random do not estimate LER accurately.
At ``Cost 1000,'' MF-LER starts sampling from the highest-fidelity function and identifying LER in more detail.
However, the predicted LER of SF-LER and SF-Random are still largely different from the truth.
At ``Cost 2000,'' SF-LER starts identifying the rough shape of the LER, and the prediction of SF-Random is not still stable.
At this cost, MF-LER provides almost precise LER estimation.
%
% By using information from the lowest fidelity functions, MF-LER only acquires points around LER from the middle and high fidelity functions.
%
The total cost 2000 is only approximately 0.05\% ($\approx 2000/3750000 \times 100$) of that used for exhaustive search on the highest-fidelity surface.
This suggests that MF-LER is effective to accelerate the search process by reducing the sampling from expensive computations.

\begin{figure}[t]
 \centering
 \begin{minipage}{0.325\hsize}
  \igr{1.}{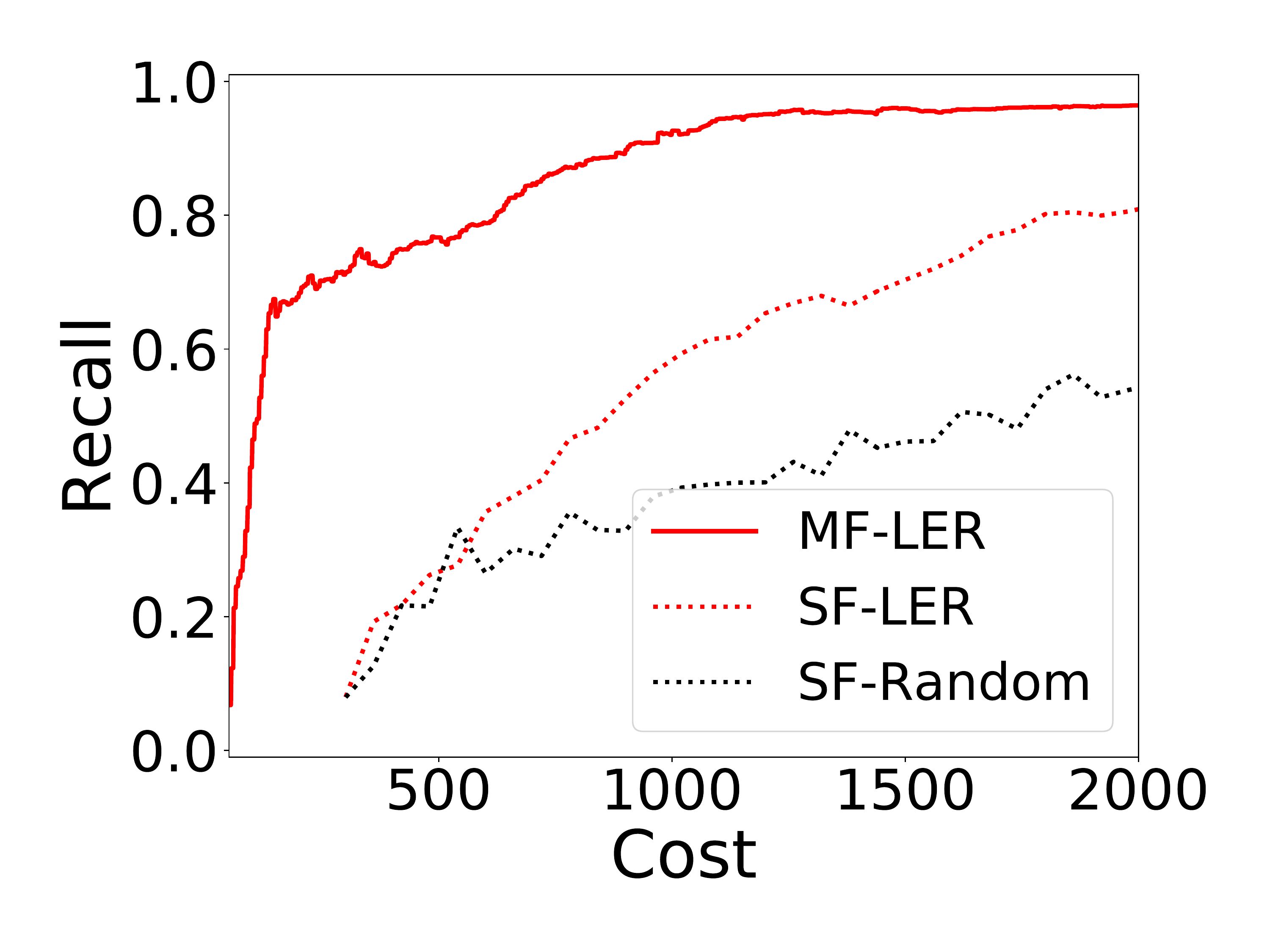}
 \end{minipage}
 \begin{minipage}{0.325\hsize}
  \igr{1.}{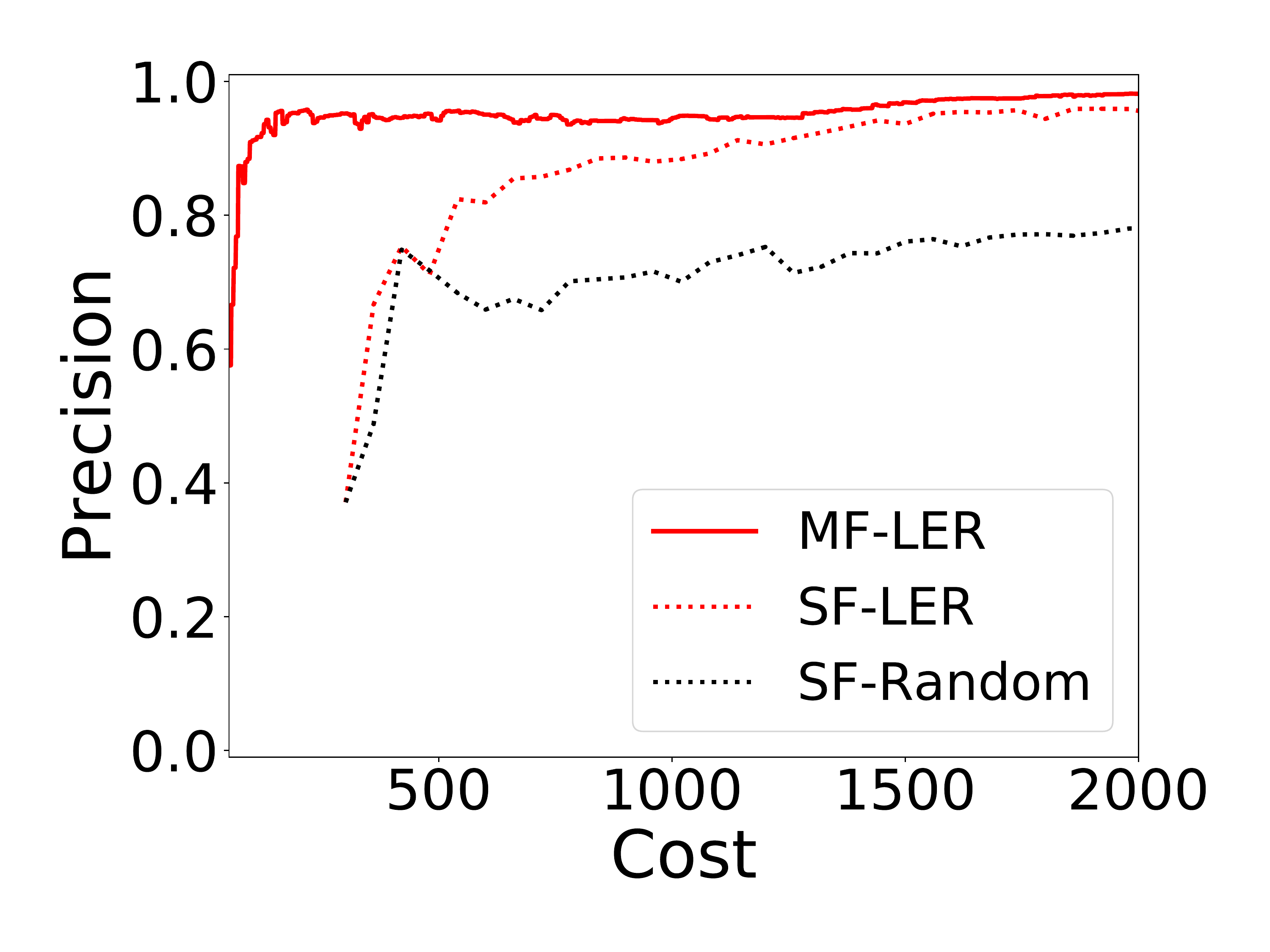}
 \end{minipage}
 \begin{minipage}{0.325\hsize}
  \igr{1.}{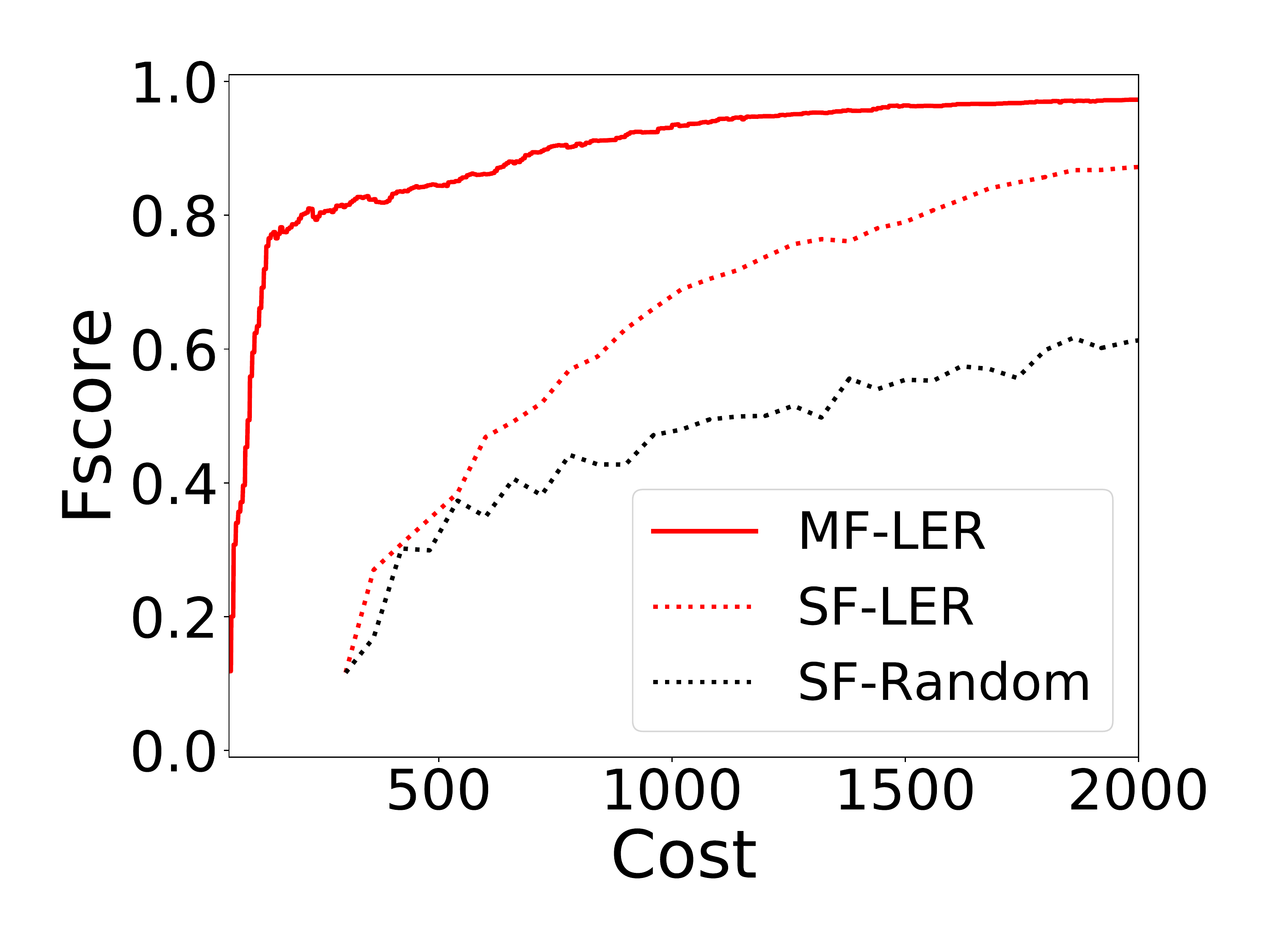}
 \end{minipage}
 \caption{
Performance evaluation with $h = 5$.
The left, middle, and right plots indicate the recall, precision, and F-score (see the main text for definition), respectively.
The horizontal axis denotes the accumulated sampling cost.
 }
 \label{fig:f-score}
\end{figure}

Figure~\ref{fig:f-score} shows the quantitative performance evaluation.
We evaluated the accuracy of LER estimation through the predicted binary label $z_{\*x}$.
Note that our objective was only to identify LER and not to approximate the entire discrepancy surface accurately that would require a higher number of samples.
We used standard evaluation measures of the classification problem called {\it{recall}}, {\it{precision}}, and {\it{F-score}}.
Because MF-LER sampled only from low-fidelity function values as initial points unlike the other two methods, the initial cost values of MF-LER in the plot are different from those of SF-LER and SF-Random.
All the results are the averages of 10 runs with random initial points.

The left plot in Fig.~\ref{fig:f-score} shows recall, defined by
\begin{align*}
 \frac{\text{The number of points } i \in LER \text{ which has } p(z_{\*x_i}=1) \geq 0.5}
 {|LER|}.
\end{align*}
This is the ratio of the number of LER points that are correctly identified over the number of points in the true LER.
This evaluates how many LER points are correctly identified.
At the beginning, recall was approximately $0.1$ for all sampling strategies owing to the absence of sampled points.
However, MF-LER rapidly increased recall substantially faster than SF-LER and SF-Random.
The middle plot in Fig.~\ref{fig:f-score} shows precision, defined by
\begin{align*}
 % \frac{\text{The number of points } i \text{ for which } p(z_{\*x_i}=1) \geq 0.5 \text{and}\ y^{(M)}_{\*x_i} \leq h}{\text{The number of points}\ i\ \text{for which}\ p(z_{\*x_i}=1) \geq 0.5}.
%  \frac{\text{The number of points } i \text{ for which } p(z_{\*x_i}=1) \geq 0.5 \text{ among } i \in LER}{\text{The number of points}\ i\ \text{for which}\ p(z_{\*x_i}=1) \geq 0.5}.
 \frac{\text{The number of points } i \in LER \text{ which has } p(z_{\*x_i}=1) \geq 0.5}{\text{The number of points } i \in \{1, \ldots, N \} \text{ which has } p(z_{\*x_i}=1) \geq 0.5}.
\end{align*}
Precision has the same numerator as recall, but the denominator is the number of points predicted as the LER.
This evaluates the specificity of prediction, which cannot be considered by recall.
The precision values were higher than the recall values in the beginning, indicating that the large part of predicted LER was actually $y^{(M)}_{\*x_i} \leq h$.
Similar to recall, MF-LER was better than SF-LER and SF-Random at all the cost values.
Because recall and precision have a tradeoff relationship, their harmonic mean, referred to as the F-score, is often used as a comprehensive evaluation criterion.
The right plot of Fig.~\ref{fig:f-score} shows the superior performance of MF-LER in terms of the F-score.

% --------------------------------------------------
% Demonstration
% --------------------------------------------------
\begin{figure}[t]
 \centering
 % \igr{1.}{demo_raster.pdf}
 \igr{1.}{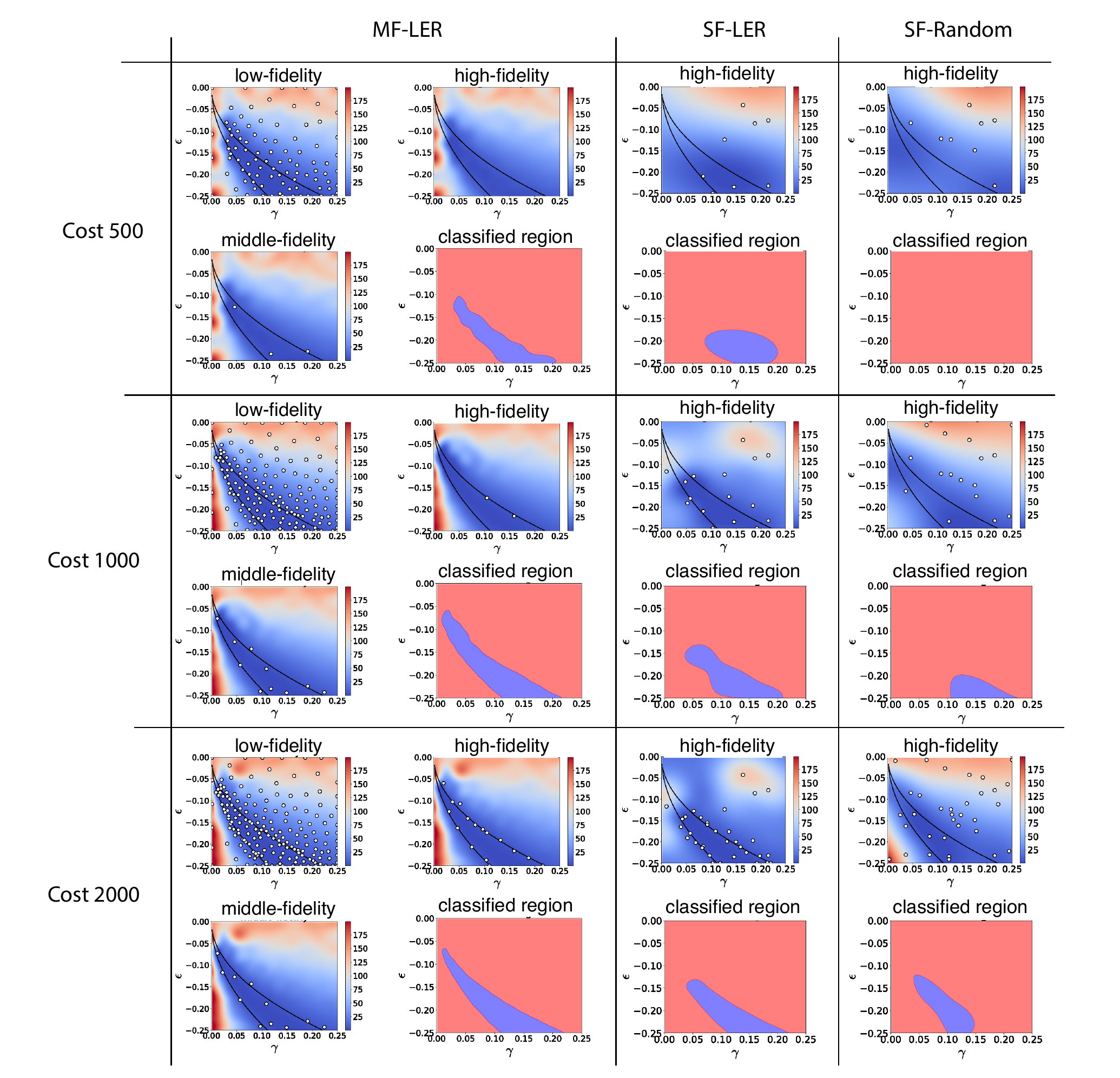}
 \caption{
 Illustrative comparison of MF-LER, SF-LER, and SF-Random.
 For each sampling strategy, results obtained using three different total sampling costs namely 500, 1000, and 2000 are shown.
 In each sampling cost of MF-LER, three heatmaps represent the predictive mean $\mu^{(m)}_{\*x}$ for $m = 1$ (upper left), $m = 2$ (lower left), and $m = 3$ (upper right).
 The right-bottom binary image in MF-LER is the predicted LER (the blue region is $p(z_{\*x_i} = 1) \geq 0.5$).
 For SF-LER and SF-Random, the heatmap for the predictive mean of the highest-fidelity function and binary image of the predicted LER are shown.
 In each heatmap, the white points represent the observed samples, and the black lines represent the boundary of the LER for $h = 5$.
 }
 \label{fig:demo}
\end{figure}

% --------------------------------------------------
% Recall/Precition/Fscore
% --------------------------------------------------

\clearpage
% --------------------------------------------------
\section{Conclusions}

We proposed an ML-based selective sampling procedure for estimating the LER of the material parameter space.
The LER is defined using the discrepancy in the precipitate shapes between the computational model and experimental image.
To efficiently explore the material parameter space, we introduced multifidelity modeling that can incorporate several levels of approximate samples.
Based on the information entropy measure, our sampling method, called MF-LER, can determine the most cost-effective pair of a sample point and a fidelity level at every iteration.
We demonstrated the effectiveness of our method by estimating the interface energy and lattice mismatch between MgZn$_2$ and $\alpha$-Mg phases in an Mg-based alloy.
The results show that lower-fidelity data are highly useful for accelerating the LER estimation drastically.
Although we focused on the Mg-based alloy as a case study, multifidelity calculations are prevalent in computational materials science, in which the computational cost often becomes a severe bottleneck.
One of our future works is to apply MF-LER to other material parameter estimation problems for efficiently analyzing a variety of materials.

\clearpage

% --------------------------------------------------
\section*{Data availability}

The data for the discrepancy surface and machine-learning code are available on request.

% --------------------------------------------------
\section*{Acknowledgment}

This work was supported by JST PRESTO (JPMJPR15NB, JPMJPR15N2 and JPMJPR16N6), Advanced Low Carbon Technology Research and Development Program (ALCA), MEXT KAKENHI (16H06538, 17H04694), and MI2I project of JST Support Program for Starting Up Innovation Hub.
%
% The authors thank Dr. Sasaki, T.T. and Dr. Hono, K. (National Institute for Materials Science, Japan) for fruitful discussion on the orientation relationship between nanometer-size precipitate and matrix phases in Mg-based alloys.

% --------------------------------------------------
\section*{Author contribution}

S.T., M.S., and M.K. constructed the machine-learning method and wrote the manuscript.
Y.T. constructed and calculated the computational model and wrote the manuscript.
H.F. constructed and calculated the computational model.
T.K. contributed to the study design.

% --------------------------------------------------
\section*{Competing interest}

The authors declare no competing interest.

\clearpage
% --------------------------------------------------
% References
% --------------------------------------------------

\bibliography{ref}

\begin{thebibliography}{10}

\bibitem{Kahlweit1975}
M.~Kahlweit.
\newblock Ostwald ripening of precipitates.
\newblock {\em Adv. Colloid Interface Sci.}, 5, 1975.

\bibitem{Ito2016}
S.~Ito, H.~Nagao, A.~Yamanaka, Y.~Tsukada, T.~Koyama, M.~Kano, and J.~Inoue.
\newblock Data assimilation for massive autonomous systems based on a
  second-order adjoint method.
\newblock {\em Phys. Rev. E}, 94, 2016.

\bibitem{Ito2017}
S.~Ito, H.~Nagao, T.~Kasuya, and J.~Inoue.
\newblock Grain growth prediction based on data assimilation by implementing
  {4DVar} on multi-phase-field model.
\newblock {\em Sci. Technol. Adv. Mater.}, 18, 2017.

\bibitem{Zhang2017}
J.~Zhang, S.~O. Poulsen, J.~W. Gibbs, P.~W. Voorhees, and H.~F. Poulsen.
\newblock Determining material parameters using phase-field simulations and
  experiments.
\newblock {\em Acta Mater.}, 129, 2017.

\bibitem{Sasaki2018}
K.~Sasaki, A.~Yamanaka, S.~Ito, and H.~Nagao.
\newblock Data assimilation for phase-field models based on the ensemble
  {Kalman} filter.
\newblock {\em Comput. Mater. Sci.}, 141, 2018.

\bibitem{Thompson1994}
M.~E. Thompson, C.~S. Su, and P.~W. Voorhees.
\newblock Equilibrium shape of a misfitting precipitate.
\newblock {\em Acta Metall. Mater.}, 42, 1994.

\bibitem{Schmidt1997}
I.~Schmidt and D.~Gross.
\newblock The equilibrium shape of an elastically inhomogeneous inclusion.
\newblock {\em J. Mech. Phys. Solids}, 45, 1997.

\bibitem{Schmidt1998}
I.~Schmidt, R.~Mueller, and D.~Gross.
\newblock The effect of elastic inhomogeneity on equilibrium and stability of a
  two particle morphology.
\newblock {\em Mech. Mater.}, 30, 1998.

\bibitem{Khachaturyan2008}
A.~G. Khachaturyan.
\newblock {\em Theory of Structural Transformations in Solids}.
\newblock Dover, 2008.

\bibitem{Porter2009}
D.~A. Porter, K.~E. Easterling, and M.~Y. Sherif.
\newblock {\em Phase Transformations in Metals and Alloys 3rd edition}.
\newblock CRC Press, 2009.

\bibitem{Clark1965}
J.~B. Clark.
\newblock Transmission electron microscopy study of age hardening in a {Mg}-5
  wt.\% {Zn} alloy.
\newblock {\em Acta Metall.}, 13, 1965.

\bibitem{Chun1969}
J.~S. Chun and J.~G. Byrne.
\newblock Precipitate strengthening mechanisms in magnesium zinc alloy single
  crystals.
\newblock {\em J. Mater. Sci.}, 4, 1969.

\bibitem{Nie1997}
J.~F. Nie and B.~C. Muddle.
\newblock Precipitation hardening of {Mg-Ca(-Zn)} alloys.
\newblock {\em Scr. Mater.}, 37, 1997.

\bibitem{Celotto2000}
S.~Celotto.
\newblock {TEM} study of continuous precipitation in {Mg}-9 wt.\% {Al}-1 wt.\%
  {Zn} alloy.
\newblock {\em Acta Mater.}, 48, 2000.

\bibitem{Smola2002}
B.~Smola, I.~Stul{\'i}kov{\'a}, F.~Buch, and B.~L. Mordike.
\newblock Structural aspects of high performance {Mg} alloys design.
\newblock {\em Mater. Sci. Eng. A}, 324, 2002.

\bibitem{Ping2003}
D.~H. Ping, K.~Hono, and J.~F. Nie.
\newblock Atom probe characterization of plate-like precipitates in a
  {Mg-RE-Zn-Zr} casting alloy.
\newblock {\em Scr. Mater.}, 48, 2003.

\bibitem{Oh2005}
J.~C. Oh, T.~Ohkubo, T.~Mukai, and K.~Hono.
\newblock {TEM} and {3DAP} characterization of an age-hardened {Mg-Ca-Zn}
  alloy.
\newblock {\em Scr. Mater.}, 53, 2005.

\bibitem{Nie2005}
J.~F. Nie, X.~Gao, and S.~M. Zhu.
\newblock Enhanced age hardening response and creep resistance of {Mg-Gd}
  alloys containing {Zn}.
\newblock {\em Scr. Mater.}, 53, 2005.

\bibitem{Sasaki2006}
T.~T. Sasaki, K.~Oh-ishi, T.~Ohkubo, and K.~Hono.
\newblock Enhanced age hardening response by the addition of {Zn} in {Mg-Sn}
  alloys.
\newblock {\em Scr. Mater.}, 55, 2006.

\bibitem{Mendis2006}
C.~L. Mendis, C.~J. Bettles, M.~A. Gibson, and C.~R. Hutchinson.
\newblock An enhanced age hardening response in {Mg-Sn} based alloys containing
  {Zn}.
\newblock {\em Mater. Sci. Eng. A}, 435-436, 2006.

\bibitem{Mendis2007}
C.~L. Mendis, K.~Oh-ishi, and K.~Hono.
\newblock Enhanced age hardening in a {Mg}-2.4 at.\% {Zn} alloy by trace
  additions of {Ag} and {Ca}.
\newblock {\em Scr. Mater.}, 57, 2007.

\bibitem{Sasaki2009}
T.~T. Sasaki, T.~Ohkubo, and K.~Hono.
\newblock Precipitation hardenable {Mg-Bi-Zn} alloys with prismatic plate
  precipitates.
\newblock {\em Scr. Mater.}, 61, 2009.

\bibitem{Oh-ishi2009}
K.~Oh-ishi, R.~Watanabe, C.~L. Mendis, and K.~Hono.
\newblock Age-hardening response of {Mg}-0.3 at.\% {Ca} alloys with different
  {Zn} contents.
\newblock {\em Mater. Sci. Eng. A}, 526, 2009.

\bibitem{Sasaki2011}
T.~T. Sasaki, K.~Oh-ishi, T.~Ohkubo, and K.~Hono.
\newblock Effect of double aging and microalloying on the age hardening
  behavior of a {Mg-Sn-Zn} alloy.
\newblock {\em Mater. Sci. Eng. A}, 530, 2011.

\bibitem{Mendis2011}
C.~L. Mendis, K.~Oh-ishi, T.~Ohkubo, and K.~Hono.
\newblock Precipitation of prismatic plates in {Mg-0.3Ca} alloys with {In}
  additions.
\newblock {\em Scr. Mater.}, 64, 2011.

\bibitem{Mendis2012}
C.~L. Mendis, K.~Oh-ishi, and K.~Hono.
\newblock Microalloying effect on the precipitation processes of {Mg-Ca}
  alloys.
\newblock {\em Metall. Mater. Trans. A}, 43, 2012.

\bibitem{Elsayed2013}
F.~R. Elsayed, T.~T. Sasaki, C.~L. Mendis, T.~Ohkubo, and K.~Hono.
\newblock Compositional optimization of {Mg-Sn-Al} alloys for higher age
  hardening response.
\newblock {\em Mater. Sci. Eng. A}, 566, 2013.

\bibitem{Bhattacharjee2013}
T.~Bhattacharjee, C.~L. Mendis, K.~Oh-ishi, T.~Ohkubo, and K.~Hono.
\newblock The effect of {Ag} and {Ca} additions on the age hardening response
  of {Mg-Zn} alloys.
\newblock {\em Mater. Sci. Eng. A}, 575, 2013.

\bibitem{Bhattacharjee2014}
T.~Bhattacharjee, T.~Nakata, T.~T. Sasaki, S.~Kamado, and K.~Hono.
\newblock Effect of microalloyed {Zr} on the extruded microstructure of
  {Mg-6.2Zn}-based alloys.
\newblock {\em Scr. Mater.}, 90-91, 2014.

\bibitem{Sasaki2015}
T.~T. Sasaki, F.~Elsayed, T.~Nakata, T.~Ohkubo, S.~Kamado, and K.~Hono.
\newblock Strong and ductile heat-treatable {Mg-Sn-Zn-Al} wrought alloys.
\newblock {\em Acta Mater.}, 99, 2015.

\bibitem{Nakata2017}
T.~Nakata, C.~Xu, R.~Ajima, K.~Shimizu, S.~Hanaki, T.~T. Sasaki, L.~Ma,
  K.~Hono, and S.~Kamado.
\newblock Strong and ductile age-hardening {Mg-Al-Ca-Mn} alloy that can be
  extruded as fast as aluminum alloys.
\newblock {\em Acta Mater.}, 130, 2017.

\bibitem{Gao2012}
Y.~Gao, H.~Liu, R.~Shi, N.~Zhou, Z.~Xu, Y.M. Zhu, J.F. Nie, and Y.~Wang.
\newblock Simulation study of precipitation in an {Mg-Y-Nd} alloy.
\newblock {\em Acta Mater.}, 60, 2012.

\bibitem{Liu2013}
H.~Liu, Y.~Gao, J.Z. Liu, Y.M. Zhu, Y.~Wang, and J.F. Nie.
\newblock A simulation study of the shape of $\beta^\prime$ precipitates in
  {Mg-Y} and {Mg-Gd} alloys.
\newblock {\em Acta Mater.}, 61, 2013.

\bibitem{Ji2014}
Y.Z. Ji, A.~Issa, T.W. Heo, J.E. Saal, C.~Wolverton, and L.-Q. Chen.
\newblock Predicting $\beta^\prime$ precipitate morphology and evolution in
  {Mg-RE} alloys using a combination of first-principles calculations and
  phase-field modeling.
\newblock {\em Acta Mater.}, 76, 2014.

\bibitem{Tsukada2014}
Y.~Tsukada, Y.~Beniya, and T.~Koyama.
\newblock Equilibrium shape of isolated precipitates in the $\alpha$-{Mg}
  phase.
\newblock {\em J. Alloy. Compd.}, 603, 2014.

\bibitem{Tsukada2019-Estimation}
Y.~Tsukada, S.~Takeno, M.~Karasuyama, H.~Fukuoka, M.~Shiga, and T.~Koyama.
\newblock Estimation of material parameters based on precipitate shape:
  efficient identification of low-error region with gaussian process modeling.
\newblock {\em Sci. Rep.}, 9:15794, 2019.

\bibitem{Pilania2017-Multi}
G.~Pilania, J.E. Gubernatis, and T.~Lookman.
\newblock Multi-fidelity machine learning models for accurate bandgap
  predictions of solids.
\newblock {\em Comput. Mater. Sci.}, 129:156 -- 163, 2017.

\bibitem{Kennedy2000-Predicting}
M.~C. Kennedy and A.~O'Hagan.
\newblock Predicting the output from a complex computer code when fast
  approximations are available.
\newblock {\em Biometrika}, 87(1):1--13, 2000.

\bibitem{MacKay2003-Information}
D.~J.~C. MacKay.
\newblock {\em Information Theory, Inference, and Learning Algorithms}.
\newblock Cambridge University Press, 2003.

\bibitem{Mura1987}
T.~Mura.
\newblock {\em Micromechanics of Defects in Solids 2nd rev. edition}.
\newblock Martinus Nijhoff, 1987.

\bibitem{Kinzoku2004}
The Japan~Institute of~Metals and Materials (ed.).
\newblock {\em Kinzoku Data Book 4th rev. edition}.
\newblock Maruzen, 2004.

\bibitem{Rasmussen2005}
C.~E. Rasmussen and C.~K.~I. Williams.
\newblock {\em Gaussian Processes for Machine Learning (Adaptive Computation
  and Machine Learning)}.
\newblock The MIT Press, 2005.

\end{thebibliography}
\bibliographystyle{unsrt}

\clearpage
% --------------------------------------------------
% Appendix
% --------------------------------------------------

\appendix

% --------------------------------------------------
\section{Computational Details of Information Entropy}
\label{sec:computation-entropy}

% The first term is entropy of Bernoulli distribution.
The first term of mutual information \eq{eq:mutual_info} is the entropy of the Bernoulli distribution.
The probability mass function of this distribution is written as
\begin{align*}
 p(z_{\*x}) &=
 \begin{cases}
  \Phi(\gamma^h_{\*x}), & \text{ if } z_{\*x} = 1\ (f^{(M)}_{\*x} \leq h), \\
  1 - \Phi(\gamma^h_{\*x}), & \text{ if } z_{\*x} = 0\ (f^{(M)}_{\*x} > h),
 \end{cases}
\end{align*}
where $\Phi(\cdot)$ is the cumulative distribution function of the standard Gaussian distribution and $\gamma^h_{\*x} = (h - \mu^{(M)}_{\*x}) / \sigma^{(M)}_{\*x}$ is the threshold $h$ normalized by the predictive mean and variance of the highest-fidelity function at $\*x$.
Hence, the first term of \eq{eq:mutual_info} is calculated easily as
\begin{align*}
 H(p(z_{\*x} | \mathcal{D}_n)) = - \Phi(\gamma^h_{\*x}) \log \Phi(\gamma^h_{\*x}) - (1 - \Phi(\gamma^h_{\*x})) \log (1 - \Phi(\gamma^h_{\*x})).
\end{align*}

The second term of \eq{eq:mutual_info} is the expectation of the entropy of the Bernoulli distribution.
%
% It is difficult to calculate the expectation of the second term analytically.
%
% Therefore, we calculate this approximately by Monte Carlo method.
We calculate this expectation using quadrature.
%
% In summary, we sample function value from the current predictive distribution and calculate the one-step ahead predictive distribution.
%
% Based on that distribution, calculate the entropy of Bernoulli distribution and take the average of each samples.
%
% Specifically, at first,
First, we draw a sample from the current predictive distribution $p(y^{(m)}_{\*x} |\*x, \cD_t)$.
%
% Let $t^{(m)}_{\*x}$ be the sampled function value of each fidelity.
%
To calculate the one-step-ahead predictive distribution of the highest-fidelity function $p(z_{\*x} | y^{(m)}_{\*x}, \mathcal{D}_n)$, we consider the two-variable marginal distribution as follows:
\begin{align*}
	\left[
	\begin{array}{c}
		y^{(m)}_{\*x} \\
		f^{(M)}_{\*x}
	\end{array}
	\right] \mid \mathcal{D}_n &\sim \mathcal{N} \left( \left[
	\begin{array}{c}
		\mu_{\*x}^{(m)} \\
		\mu^{(M)}_{\*x}
	\end{array}
	\right], \left[
	\begin{array}{cc}
		\sigma^{2(m)}_{\*x} + \sigma^2_{\rm noise} & \sigma^{2(mM)}_{\*x} \\
		\sigma^{2(mM)}_{\*x} & \sigma^{2(M)}_{\*x}
	\end{array}
	\right] \right),
\end{align*}
where $\sigma^{2(mM)}_{\*x}$ is the covariance between $y^{(m)}_{\*x}$ and $f^{(M)}_{\*x}$.
%
% By property of conditional gaussian distribution, one-step ahead predictive distritubion is obtained as
Through the conditional Gaussian distribution, we obtain
\begin{align*}
	f^{(M)}_{\*x} \mid y^{(m)}_{\*x}, \mathcal{D}_n \sim \mathcal{N} \bigl(\mu^{(M)}_{\*x \mid y^{(m)}_{\*x}}, \sigma^{(M)}_{\*x \mid y^{(m)}_{\*x}} \bigl),
\end{align*}
where,
\begin{align*}
	\mu^{(M)}_{\*x \mid y^{(m)}_{\*x}} &= \frac{\sigma^{2(mM)}_{\*x} (y^{(m)}_{\*x} - \mu^{(m)}_{\*x})}{\sigma^{2(m)}_{\*x} + \sigma^2_{\rm noise}} + \mu^{(M)}_{\*x}, \\
	\sigma^{(M)}_{\*x \mid y^{(m)}_{\*x}} &= \sigma^{2(M)}_{\*x} - \frac{ \bigl( \sigma^{2(mM)}_{\*x}\bigl)^2}{\sigma^{2(m)}_{\*x} + \sigma^2_{\rm noise}}.
\end{align*}
%
% Then, the entropy based on one sample is calculated as
Then, the entropy of $p(z_{\*x} | y^{(m)}_{\*x}, \mathcal{D}_n)$ is written as
\begin{align*}
	H(p(z_{\*x} | y^{(m)}_{\*x}, \mathcal{D}_n)) &= - \Phi(\Gamma^h_{\*x}) \log \Phi(\Gamma^h_{\*x}) - (1 - \Phi(\Gamma^h_{\*x})) \log (1 - \Phi(\Gamma^h_{\*x})),
\end{align*}
where $\Gamma^h_{\*x} = (h - \mu^{(M)}_{\*x \mid y^{(m)}_{\*x}}) / \sigma^{(M)}_{\*x \mid y^{(m)}_{\*x}}$ is the threshold normalized by the one-step-ahead predictive mean and variance of the highest-fidelity function at $\*x$.
Finally, we approximate the expectation
$\mathbb{E}_{p(y^{(m)}_{\*x} | \mathcal{D}_n)} \bigl[ H(p(z_{\*x} | y^{(m)}_{\*x}, \mathcal{D}_n)) \bigl]$
using quadrature.
% Finally, we approximate the expectation by Monte Carlo method
% \begin{align*}
% 	\mathbb{E}_{p(y^{(m)}_{\*x} | \mathcal{D}_n)}
% 	\bigl[
% 	H(p(z_{\*x} | y^{(m)}_{\*x}, \mathcal{D}_n))
% 	\bigl] &=
% 	\frac{1}{K} \sum_{y^{(m)}_{\*x} \in \mathcal{Y}} H(p(z_{\*x} | y^{(m)}_{\*x}, \mathcal{D}_n)),
% \end{align*}
% where $\mathcal{Y}$ is a set of sampled function values and $K$ is the number of samples.
%
% In our experiment, the number of samples is set as $K = 1000$.

% --------------------------------------------------
\section{Settings of Gaussian processes}
\label{sec:settings}

% In practical, Gaussian process regression is often used with normalization by observations and we follow this.
%
We used the Gaussian kernel for both kernels $k_1(\*x_i, \*x_j)$ and $k_g(\*x_i, \*x_j)$, which are defined as
\begin{align*}
	k_1(\*x_i, \*x_j) &= \sigma^2_f \exp \biggl( - \frac{\|\*x_i - \*x_j\|^2 }{2 \ell^2_f}\biggl), \\
	k_g(\*x_i, \*x_j) &= \sigma^2_g \exp \biggl( - \frac{\|\*x_i - \*x_j\|^2 }{2 \ell^2_g}\biggl),
\end{align*}
where $\sigma^2_g, \ell^2_g, \sigma^2_f$, and $\ell^2_f$ are hyperparameters.
The variance of noise term $\sigma^2_{\rm noise}$ is also a hyperparameter.
We set $\sigma^2_{\rm noise} = 10^{-8}, \sigma^2_g = 10^{-2}, \ell^2_g = 10, \sigma^2_f=1$.
$\ell^2_f$ is optimized by {\it marginal likelihood maximization} \cite{Rasmussen2005} per $5$ iteration.

% % --------------------------------------------------
% \section{Experiment with $h=1$}
% \label{sec:h=1}

% Figure~\ref{fig:f-score_h=1} shows the experimental results when we set the thresholed $h = 1$.
% %
% % The proposed method identified the LER accurately after using $2000$ costs despite the region was narrower than that of $h = 5$.
% The proposed method identified the LER accurately after the total cost amounts to $2000$ despite the region was narrower than that of $h = 5$.
% % --------------------------------------------------
% % Recall/Precition/Fscore
% % --------------------------------------------------
% \begin{figure}[t]
% 	\centering
% 	\begin{minipage}{0.325\hsize}
% 		\igr{1.}{Real-Recall-h=1.pdf}
% 	\end{minipage}
% 	\begin{minipage}{0.325\hsize}
% 		\igr{1.}{Real-Precision-h=1.pdf}
% 	\end{minipage}
% 	\begin{minipage}{0.325\hsize}
% 		\igr{1.}{Real-Fscore-h=1.pdf}
% 	\end{minipage}
% 	\caption{
% 		Performance evaluation with $h = 1$.
% 		%
% 		The image on the left is the ratio of points $i$ for which $p(z_{\*x_i} = 1) \geq 0.5$ among the points in LER (recall).
% 		%
% 		The image at the middle is the ratio of points $i$ for which $y^{(M)}_{\*x_i} \leq h$ among the points in the predicted LER $p(z_{\*x_i} = 1) < 0.5$ (precision).
% 		%
% 		The image on the right is the harmonic mean of recall and precision (F-score).
% 		}
%  \label{fig:f-score_h=1}
% \end{figure}

\end{document}